\documentclass[conference]{IEEEtran}
\usepackage{amsfonts}
\usepackage{graphicx}
\usepackage{epsfig} 
\usepackage{epstopdf}
\usepackage{amsmath}
\usepackage{subfigure}
\usepackage{multirow}
\usepackage{makecell}
\usepackage{booktabs}
\usepackage{algorithm}
\usepackage{algorithmic}
\usepackage{amsthm}
\usepackage{bm}

\usepackage[square, comma, sort&compress, numbers]{natbib}

\usepackage{tikz} 
\usepackage{pgfplots} 
\usepackage{pgfplotstable} 
\usepackage{textcomp}

\pgfplotsset{grid style={step=1.0, dotted,black}} 
\pgfplotsset{compat=newest} 

\definecolor{mycolor1}{RGB}{229,0,0}
\definecolor{mycolor2}{RGB}{21,176,26}
\definecolor{mycolor3}{RGB}{5,4,170}

\begin{document}
	
	\title{Joint Neural Network Equalizer and Decoder}
	
	\author{\IEEEauthorblockN{Weihong Xu$^{1,2,3}$, Zhiwei Zhong$^{1,2,3}$, Yair Be'ery$^4$, Xiaohu You$^{2}$, and Chuan Zhang$^{1,2,3,*}$}
		\IEEEauthorblockA{$^{1}$Lab of Efficient Architecture for Digital Communication and Signal Processing (LEADS)\\
			$^{2}$National Mobile Communications Research Laboratory\\
            $^{3}$Quantum Information Center, Southeast University, China\\
			$^{4}$School of Electrical Engineering, Tel-Aviv University, Israel\\
			Email: $^2$\{wh.xu, zwzhong, xhyu, chzhang\}@seu.edu.cn, $^4$ybeery@eng.tau.ac.il}
	}
	
	\maketitle
	
	\begin{abstract}
		Recently, deep learning methods have shown significant improvements in communication systems. In this paper, we study the equalization problem over the nonlinear channel using neural networks. The joint equalizer and decoder based on neural networks are proposed to realize blind equalization and decoding process without the knowledge of channel state information (CSI). Different from previous methods, we use two neural networks instead of one. First, convolutional neural network (CNN) is used to adaptively recover the transmitted signal from channel impairment and nonlinear distortions. Then the deep neural network decoder (NND) decodes the detected signal from CNN equalizer. Under various channel conditions, the experiment results demonstrate that the proposed CNN equalizer achieves better performance than other solutions based on machine learning methods. The proposed model reduces about $2/3$ of the parameters compared to state-of-the-art counterparts. Besides, our model can be easily applied to long sequence with $\mathcal{O}(n)$ complexity.
	\end{abstract}


	\section{Introduction}\label{sec:Introduction}
		
	Recently, deep learning-aided communication systems have demonstrated amazing improvements regarding performance and adaptability. An end-to-end communications systems based on deep learning is introduced in \cite{dorner2018deep}. For channel decoding, the authors in \cite{nachmani2018deep} propose a systematic framework to construct deep neural network decoder of BCH code, which achieves encouraging performance. The neural network decoder (NND) of polar codes \cite{arikan2009channel} in \cite{gruber2017deep,xu2017improved} shows significant improvement compared with conventional methods.
	
	Except for noise, communication channel usually introduces inter-symbol interference (ISI) to the transmitted signal. The amplifiers and mixers will additionally cause nonlinear distortion. Channel equalization is exploited to deal with these dispersive effects and delivers recovered signal to the channel decoder. Then the channel decoder performs decoding to correct errors during transmission.
	
	Various machine learning approaches have been utilized to address the nonlinear equalization without accurate knowledge of channel state information (CSI). These methods include deep neural network (DNN) \cite{ye2017initial}, convolutional neural network (CNN) \cite{caciularu2018blind}, Gaussian processes for classification (GPC) \cite{olmos2010joint} and support vector machine (SVM) \cite{sebald2000support}, enabling the receiver side adaptively realize equalization. However, DNN equalizer in \cite{ye2017initial} is only practical for short codes and requires large amounts of parameters. The equalizers in \cite{salamanca2010channel,olmos2010joint} both use multiple training sequences to estimate the channel filter coefficients of ISI. These two methods also require other a priori knowledge such as the distribution of channel filter coefficients and the Gaussian noise variance. These assumptions may not hold in practice. Consequently, it is hard to obtain the accurate results of CSI analytically without the a priori information. As indicated in \cite{olmos2010joint}, the performance of SVM equalizer after decoder is poor.
	
	In this work, we propose neural network based joint channel equalizer and decoder. The CNN equalizer is first utilized to compensate the distortion of signal. Then the other polar NND is cascaded after CNN to decode the recovered sequence. The networks are trained with an offline and end-to-end manner, which is capable of carrying out equalization and decoding without the side information of CSI. The experiments show that promising bit error rate (BER) improvement is achieved compared with the other machine learning methods.
	
	The rest of this paper is organized as follows. Brief preliminaries about system setup and ML equalizer are reviewed in Section \ref{sec:Preliminaries}. In Section \ref{sec:proposed}, the details of proposed joint neural network equalizer and decoder as well as corresponding training methods are presented. The experiment results are shown in Section \ref{sec:Experiment_results} and analysis is given in Section \ref{sec:Analysis}. Section \ref{sec:Conclusion} finally concludes the paper.

	\section{Preliminaries}\label{sec:Preliminaries}
	
	\begin{figure*}[ht]
		\centering
		\includegraphics[width=\linewidth]{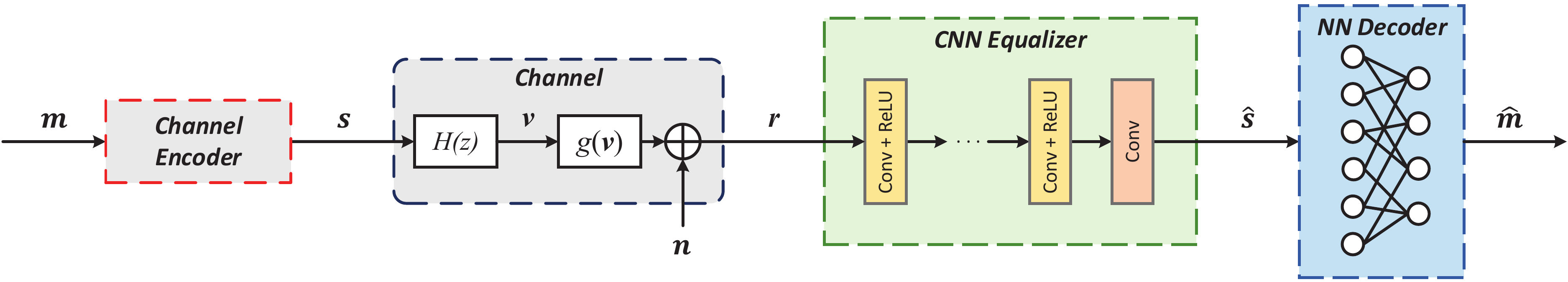}
		\caption{System Architecture.}
		\label{fig:system_arch}
	\end{figure*}

%
%
	
	\subsection{System Setup}

	Throughout this paper, the discrete-time dispersive digital-communication system with a nonlinear channel is considered. In this case, the ISI, nonlinearities, and AWGN jointly result in the channel distortion. The overall system architecture is depicted in Fig. \ref{fig:system_arch}.
	
	\subsubsection{Inter-symbol Interference}
	The channel with ISI is modeled as a finite impulse response (FIR) filter. The corresponding filter coefficients with length $L$ are given by $\bm{h} = [h_{1},h_{2},...,h_{L}]^{T}$. The signal with ISI is equivalent to the convolution of channel input with the FIR filter as follows:
	\begin{equation}\label{eq:isi_conv}
		\bm{v} = \bm{s} \otimes \bm{h},
	\end{equation}
	where $\bm{s}$ denotes the channel input sequence after modulation and $\otimes$ represents linear convolution operation.

	\subsubsection{Nonlinear Distortion}
	The nonlinearities in the communication system are mainly caused by amplifiers and mixers, which can be modeled by a nonlinear function $g[\cdot]$. After being transmitted through the channel, the output $\bm{r}$ involved with ISI, nonlinearities, and AWGN is expressed as:
	\begin{equation}\label{eq:channel_nl}
		r_{i} = g[v_{i}] + n_{i}.
	\end{equation}

	\subsection{Maximum Likelihood Equalizer}
	
	For the ML channel equalizer, a training sequence $\bm{s}^{\circ} = \{s_{1}^{\circ}, s_{2}^{\circ}, ..., s_{n}^{\circ}\}$ with $n$ known bits is sent to the receiver side to estimate the channel condition. The ML criterion for the channel coefficients is considered:
	\begin{equation}\label{eq:h_ML}
		\hat{\bm{h}}_{ML} = \arg\max_{\bm{h}} p(\bm{r}^{\circ} | \bm{s}^{\circ}, \bm{h}).
	\end{equation}
	
	After the channel coefficients are estimated, the following posterior probability of each transmitted bit can be calculated:
	\begin{equation}\label{eq:prob_bit}
		p(s_{i}=s | \bm{r}, \hat{\bm{h}}_{ML}), i =1,2...N.
	\end{equation}
	
	The output of ML equalizer is then fed into the channel decoder to obtain the estimate $\hat{\bm{m}}$ for information bits $\bm{m}$.
	
	\section{Joint Neural Network Equalizer and Decoder}\label{sec:proposed}
	
	\subsection{Convolutional Neural Network Equalizer}
	
	
	Due to the nonlinear nature of distortion, various neural networks are exploited to realize adaptive equalization. DNN \cite{ye2017initial} show promising performance. However, the densely connected networks require a large number of parameters and the complexity grows exponentially with the code length. Although recurrent neural network (RNN) \cite{kechriotis1994using} can process long code, some performance degradation appears since the neural network is unable to make the best use of the continuous inputs. Compared to other machine learning based equalization schemes, CNN requires minimal data preprocessing and have strong capabilities of feature extraction. A collection of neurons in CNN only respond to a restricted area of given inputs, which is suitable for aforementioned nonlinear channel model since the ISI only exists between consecutive bits of transmitted sequence and the influence of nonlinear distortion is independent for each bit.
		
	The architecture of proposed convolutional neural network equalizer is illustrated in Fig. \ref{fig:system_arch}. The adopted CNN is composed of several convolutional layers with rectified linear units (ReLU) while the last layer only contains convolution operation. Under the circumstances of channel equalization, the input of CNN is 1-D real vector instead of typical 2-D image in the field of computer vision. Hence, we rewrite the 2-D convolution with ReLU into 1-D form:
	\begin{equation}\label{eq:cnn_eq_1d_conv}
		\mathbf{y}_{i,j}= \sigma( \sum_{c=1}^{C}\sum_{k=1}^{K} \mathbf{W}_{i,c,k}\mathbf{x}_{c,k+j} + b_{i}),
	\end{equation}
	where $\mathbf{W}\in \mathbb{R}^{M\times C\times K}$ denotes the weights tensor of $M$ filters with $C$ channels in a layer, each containing $1\times K$ sized filter. Besides, $b_{i}$ is the $i$-th element of the bias vector $\bm{b}\in \mathbb{R}^{M}$ and $\sigma(\cdot)$ here denotes the ReLU function $\max(x, 0)$.
	
	Different from ML equalizers \cite{salamanca2010channel}, proposed convolutional neural network equalizer is trained with an offline and end-to-end manner. The received training channel output $\bm{r}^{\circ}$ and true channel input $\bm{s}^{\circ}$ are required during training. The optimal parameters $\hat{\bm{\theta}}=\{\mathbf{W}, \bm{b}\}$ of neural network are estimated instead of the ML criterion $\hat{\bm{h}}_{ML}$:
	\begin{equation}\label{eq:theta_ML}
		\hat{\bm{\theta}} = \arg\max_{\bm{\theta}} p(\hat{\bm{s}}=\bm{s}^{\circ} | \bm{r}^{\circ}, \bm{\theta}).
	\end{equation}
	
	The optimal parameters $\hat{\bm{\theta}}$ can be obtained by using the stochastic gradient descent (SGD) with back propagation. Once the neural network has been trained, only channel output $\bm{r}$ is fed into the neural network equalizer for inference phase, which provides a blind equalization.
	
	\subsection{Deep Neural Network Decoder}
	Deep neural network of \cite{gruber2017deep} shares similar feedforward structure with CNN, but the neurons in a fully-connected layer are densely connected to the previous layer and each neuron associates with its weight. According to \cite{gruber2017deep}, deep neural network can be trained to achieve optimal BER performance over short code length. We use a multiple-layer NND to decode the output of convolutional neural network equalizer (see Fig. \ref{fig:system_arch}). The input of NND is a 1-D vector from CNN equalizer. The computation of single layer in NND can be formulated as following matrix multiplication:
	\begin{equation}\label{eq:dnn}
		\mathbf{y} = \sigma(\mathbf{W} \mathbf{x} + \mathbf{b}),
	\end{equation}
	where $\mathbf{W}$ is the weights corresponding the input $\mathbf{x}$. Note that $\sigma(\cdot)$ denotes the activation function (ReLU or sigmoid). The sigmoid activation $(1+\exp^{-x})^{-1}$ is applied to squash the output of NND to range $(0,1)$.
	
	
	\subsection{Training}
	The performance of neural networks greatly depends on the training process. First, the loss function should be carefully selected to give an accurate measure of the distance between neural network outputs and correct labels. Then the hyperparameters related to network structure and training also determine the capabilities of neural network.
	
	Proposed neural network equalizer and decoder consist of two neural networks. For the CNN equalizer, mean squared error (MSE) is adopted as the equalization loss function:
	\begin{equation}\label{eq:loss_eq}
		\mathcal{L}(\hat{\bm{s}}, \bm{s}) = \dfrac{1}{N} \sum_{i} | \hat{s}_{i} - s_{i} |^{2},
	\end{equation}
	where $\hat{\bm{s}}$ denotes the estimated equalization output of CNN equalizer and $\mathbf{s}$ is the actual transmitted signal.
	
	For the NND, we use binary cross-entropy (BCE) as the decoding loss function:
	\begin{equation}\label{eq:loss_nnd}
		\mathcal{L}(\hat{\bm{m}}, \bm{m}) = \dfrac{1}{N} \sum_{i} \hat{m}_{i} \log(m_{i}) + (1-\hat{m}_{i})\log(1-m_{i}),
	\end{equation}
	where $\hat{\bm{m}}$ denotes the decoded output of NND and $\bm{m}$ is the original information sequence.
	
	Hence, the total loss of these two neural networks becomes:
	\begin{equation}\label{eq:total_loss}
		\mathcal{L}_{total} = \mathcal{L}(\hat{\bm{s}}, \bm{s}) + \mathcal{L}(\hat{\bm{m}}, \bm{m}).
	\end{equation}
	
	The goal of training is to minimize the total loss of Eq. (\ref{eq:total_loss}). SGD and other deep learning approaches can be utilized to find to optimal network parameters resulting in the minimal loss.

	\section{Experiments}\label{sec:Experiment_results}
	
	\subsection{Experimental Setup}
	The proposed models are implemented on the advanced deep learning framework \textit{PyTorch}. GPU (NVIDIA GeForce GTX 1080 Ti) is used to accelerate training. Polar code with $1/2$ rate and binary phase shift keying (BPSK) modulation is adopted. Without loss of generality, we use the following notations to define the architecture of neural networks used in this paper. The CNN with $D$ convolutional layers, the $i$-th layer containing $M_{i}$ filters with size $1\times K$, is denoted as CNN $\{M_{1}, M_{2}, ..., M_{D}\}$. The deep neural network with $D$ layers, the $i$-th layer containing $H_{i}$ nodes, is expressed by DNN $\{H_{1}, H_{2}, ..., H_{D}\}$. It should be noted that the first and last layer of DNN are the input layer and the output layer, respectively.
	
	\begin{table}[ht]
		\centering
		\caption{Summary of experimental settings.}
		\label{tab:experiment_set}
		\begin{tabular}{c|c}
			\toprule
			\textbf{Parameters} & \textbf{Value}\\
			\midrule
				CNN structure & $\{6, 12, 24, 12, 6, 1\}, K=3$\\
				DNN structure & $\{16, 128, 64, 32, 8\}$\\
				Channel & AWGN \\
				SNR Range & 0 dB to 11 dB \\
				Training Samples per SNR & 20 \\
				Testing Samples per SNR & 50000 \\
				Mini-batch Size & 240 \\
				Optimizer     	& Mini-batch SGD with Adam \\
				Learning Rate 	& Lr=0.001 \\
				Weights Initialization & $\mathcal{N}\sim(\mu=0, \sigma=1)$\\
			\bottomrule
		\end{tabular}
	\end{table}
	
	We adopt the structure of CNN $\{6, 12, 24, 12, 6, 1\}$ with padding size $1$ since $1\times 3$ filter size can not only maintain the length of input vector but also preserve the dimension of intermediate feature maps. Besides, this configuration balances complexity and performance. According to our test, no more gain is observed over larger CNN while fewer filters will cause performance degradation. DNN structure is $\{16, 128, 64, 32, 8\}$, which has been demonstrated to achieve optimal decoding performance for short polar codes in \cite{gruber2017deep}. 	
	
	The training and testing data are both randomly generated over AWGN channel with ISI and nonlinearity distortion from SNR 0 dB to 11 dB. Each training mini-batch contains 240 samples (20 frames per SNR). 50000 frames per SNR are used for testing. The length of each frame is 8-bit. We use mini-batch SGD with back propagation to update the weights. Learning rate is set to 0.001 and Adam optimizer \cite{kingma2014adam} is utilized to adjust the learning rate dynamically.The weights are initialized with normal distribution $\mathcal{N}(\mu=0, \sigma=1)$. CNN is trained for 5000 iterations. The experimental settings are summarized in Table \ref{tab:experiment_set}.
	
	\subsection{Experiment Results over Linear Channel}
	In the first experiment, the dispersive channel with ISI and AWGN. To remain consistent with \cite{kechriotis1994using,salamanca2010channel,olmos2010joint}, the associated impulse response with length $L=3$ can be expressed as the following equation:
	\begin{equation}\label{eq:ISI_H(z)}
		H(z) = 0.3482 + 8704z^{-1} + 0.3482z^{-2}.
	\end{equation}
	
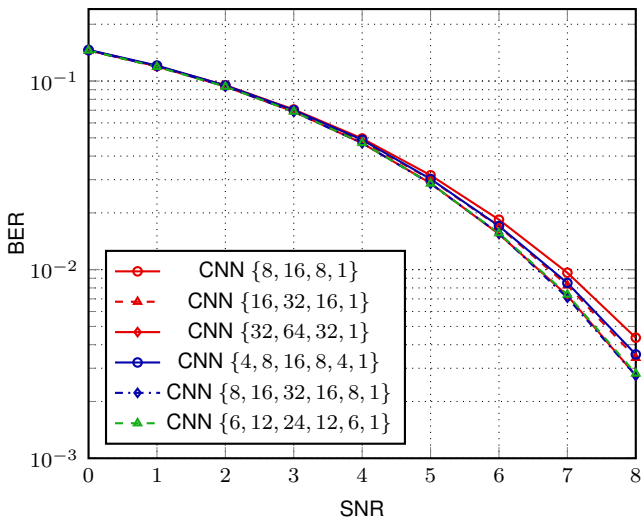
\begin{figure}[!h]
		\centering
		\begin{tikzpicture}[font=\sffamily\footnotesize]		
		\begin{semilogyaxis}[
		height=0.32\textheight,
		width=\linewidth,
		xmin=0, xmax=8,
		ymin=1e-3,
		grid=both,
		xtick = {0,1,...,8},
		xlabel=SNR,
		ylabel=BER,
		y label style={at={(axis description cs:-0.1,.5)}},
		line width=0.8pt,
		legend pos=south west
		]

		\addplot[mycolor1, mark=o, mark options={scale=0.8, solid}] coordinates {
			(0, 0.14558187 )
			(1, 0.12002031 )
			(2, 0.09501187 )
			(3, 0.07066062 )
			(4, 0.04961562 )
			(5, 0.03166281 )
			(6, 0.01841375 )
			(7, 0.00963375 )
			(8, 0.00435625 )
		};
		\addlegendentry{CNN $\{8,16,8,1\}$}
		
		\addplot[mycolor1, dashed, mark=triangle, mark options={scale=0.8, solid}] coordinates {
			(0, 0.14513906 )
			(1, 0.11906687 )
			(2, 0.09416625 )
			(3, 0.06979093 )
			(4, 0.04795718 )
			(5, 0.03008031 )
			(6,	0.01689062 )
			(7, 0.00825562 )
			(8, 0.00340875 )
		};
		\addlegendentry{CNN $\{16,32,16,1\}$}
		
		\addplot[mycolor1, mark=diamond, mark options={scale=0.8, solid}] coordinates {
			(0, 0.14575406 )
			(1, 0.11979312 )
			(2, 0.09354593 )
			(3, 0.06884937 )
			(4, 0.04684031 )
			(5, 0.02870781 )
			(6, 0.01544843 )
			(7, 0.00728812 )
			(8, 0.00274593 )
		};
		\addlegendentry{CNN $\{32,64,32,1\}$}

		\addplot[mycolor3, mark=o, mark options={scale=0.8, solid}] coordinates {
			(0, 0.14626468 )
			(1, 0.12081906 )
			(2, 0.09471937 )
			(3, 0.07014406 )
			(4, 0.04877375 )
			(5, 0.03024843 )
			(6, 0.01703656 )
			(7, 0.00849843 )
			(8, 0.00355312 )
		};
		\addlegendentry{CNN $\{4,8,16,8,4,1\}$}
		
		\addplot[mycolor3, dash dot, mark=diamond, mark options={scale=0.8, solid}] coordinates {
			(0, 0.14503375 )
			(1, 0.11980812 )
			(2, 0.09365468 )
			(3, 0.06871843 )
			(4, 0.04688343 )
			(5, 0.02857750 )
			(6, 0.01553062 )
			(7, 0.00709906 )
			(8, 0.00274187 )
		};
		\addlegendentry{CNN $\{8,16,32,16,8,1\}$}
		
		\addplot[mycolor2, dashed, mark=triangle, mark options={scale=0.8, solid}] coordinates {
			(0, 0.14527187 )
			(1, 0.11902562 )
			(2, 0.09339250 )
			(3, 0.06888531 )
			(4, 0.04696031 )
			(5, 0.02861562 )
			(6, 0.01561218 )
			(7, 0.00737781 )
			(8, 0.00281406 )
		};
		\addlegendentry{CNN $\{6,12,24,12,6,1\}$}
		
		\end{semilogyaxis}
		\end{tikzpicture}
		\caption{BER comparison of CNN equalizer with different structures over linear channel.}
		\label{fig:cnn_size_vs_performance}
	\end{figure}

	There is no strict rules that precisely measure the proper size of neural networks for a specific task. Hence, various configurations of CNN are tested to determine the best structure. The BER results for each configuration is shown in Fig. \ref{fig:cnn_size_vs_performance}. It can be observed that more filters guarantee better performance under $4$-layer CNN (\textcolor{mycolor1}{red} line). However, deeper structure outperforms the shallow ones with equal or less amount of parameters (\textcolor{mycolor3}{blue} and \textcolor{mycolor2}{green} line). CNN $\{6,12,24,12,6,1\}$ (containing 2257 weights) achieves very close performance with CNN $\{8,16,32,16,8,1\}$ (containing 3969 weights) and CNN $\{32,64,32,1\}$ (containing 12609 weights).
	
\begin{figure}[!h]
		\centering
		\begin{tikzpicture}[font=\sffamily\footnotesize]
		\begin{semilogyaxis}[
		height=0.32\textheight,
		width=\linewidth,
		xmin=0, xmax=8,
		ymin=1e-3,
		grid=both,
		xtick = {0,1,...,8},
		xlabel=SNR,
		ylabel=BER,
		y label style={at={(axis description cs:-0.1,.5)}},
		line width=0.8pt,
		legend pos=south west,
		legend entries=
		{Perfect CSI,
			ML-BCJR,
			Bayesian,
			CNN,
			$n=10$,
			$n=20$}
		]
		\addlegendimage{no markers, dash dot}
		\addlegendimage{mycolor2, no markers, dashed}
		\addlegendimage{mycolor3, no markers}
		\addlegendimage{mycolor1, mark=diamond}
		\addlegendimage{only marks, mark=o}
		\addlegendimage{only marks, mark=triangle}

		\addplot[dash dot, mark=none] coordinates {
			(0, 0.14314 )
			(1, 0.11786 )
			(2, 0.09275 )
			(3, 0.06765 )
			(4, 0.04511 )
			(5, 0.02656 )
			(6, 0.01329 )
			(7, 0.00542 )
			(8, 0.00179 )
		};
		
		\addplot[mycolor2, dashed, mark=o, mark options={scale=0.8, solid}] coordinates {
			(0, 0.19762 )
			(1, 0.17267 )
			(2, 0.14596 )
			(3, 0.11735 )
			(4, 0.08732 )
			(5, 0.060452 )
			(6, 0.037744 )
			(7, 0.021036 )
			(8, 0.009967 )
		};
		
		\addplot[mycolor3, mark=o, mark options={scale=0.8, solid}] coordinates {
			(0, 0.19161 )
			(1, 0.16777 )
			(2, 0.14076 )
			(3, 0.11377 )
			(4, 0.084257 )
			(5, 0.057862 )
			(6, 0.035159 )
			(7, 0.018857 )
			(8, 0.008719 )
		};
		
		\addplot[mycolor2, dashed, mark=triangle, mark options={scale=0.8, solid}] coordinates {
			(0, 0.17314 )
			(1, 0.14908 )
			(2, 0.12083 )
			(3, 0.09152 )
			(4, 0.06421 )
			(5, 0.04031 )
			(6, 0.02145 )
			(7, 0.00967 )
			(8, 0.00357 )
		};

		\addplot[mycolor3, mark=triangle, mark options={scale=0.8, solid}] coordinates {
			(0, 0.16968 )
			(1, 0.14610 )
			(2, 0.11842 )
			(3, 0.08969 )
			(4, 0.06293 )
			(5, 0.03951 )
			(6, 0.02103 )
			(7, 0.00948 )
			(8, 0.00350 )
		};
		
		\addplot[mycolor1, mark=diamond] coordinates {
			(0, 0.14503375 )
			(1, 0.11980812 )
			(2, 0.09365468 )
			(3, 0.06871843 )
			(4, 0.04688343 )
			(5, 0.02857750 )
			(6, 0.01553062 )
			(7, 0.00709906 )
			(8, 0.00274187 )
		};
		\end{semilogyaxis}
		\end{tikzpicture}
		\caption{BER comparison of proposed CNN, Bayesian \cite{salamanca2010channel} and ML-BCJR equalizers over linear channel.}
		\label{fig:ber_linear_channel}
	\end{figure}
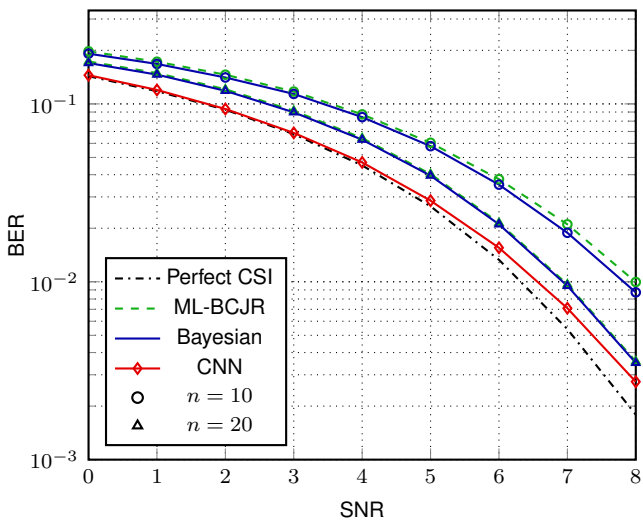

	The comparison between proposed CNN equalizer and other equalization schemes is also conducted (see Fig. \ref{fig:ber_linear_channel}). The structure of CNN equalizer is $\{6,12,24,12,6,1\}$. ML-BCJR and Bayesian methods \cite{salamanca2010channel} are tested with different lengths of the training sequence (10 and 20). Besides, the performance of ML equalizer with the knowledge of perfect CSI is also depicted. Proposed CNN equalizer approaches to perfect CSI under low SNR range and outperforms both ML-BCJR and Bayesian equalizer by about 0.5 dB.
	
	\subsection{Experiment Results over Nonlinear Channel}
\begin{figure}[ht]
		\centering
		\includegraphics[width=0.85\linewidth]{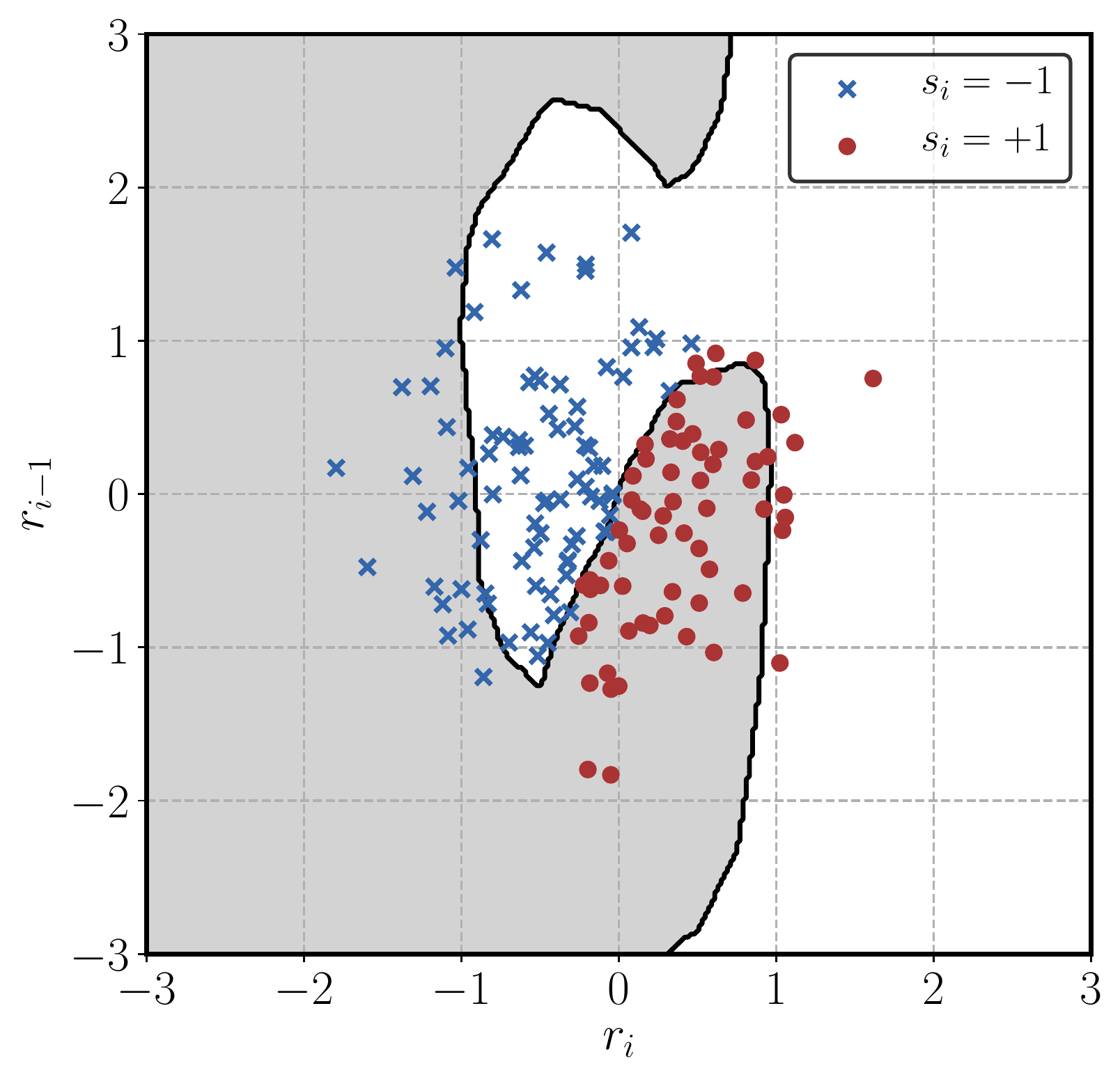}
		\caption{Decision boundary of proposed CNN equalizer with nonlinear channel $\bm{h}= [1, 0.5]$, $g(v) = v -0.9 v^{3}$ and $\text{SNR} = 1$.}
		\label{fig:figure_decision_bound}
	\end{figure}

	In the second experiment, the nonlinearity is additionally considered. To intuitively illustrate the nonlinear function of CNN, the decision boundary of proposed equalizer is drawn in Fig. \ref{fig:figure_decision_bound}. The adopted CNN structure is $\{6,12,24,12,6,1\}$ and the corresponding nonlinear channel is $\bm{h}= [1, 0.5]$ and $g(v) = v -0.9 v^{3}$. The shaded region represents that the hard decision of CNN output is $\hat{s}_{i} = +1$ when $s_{i} = +1$ transmitted. It can be seen that CNN constructs a nonlinear decision boundary to recover the original symbol from the received signal mixed with ISI and nonlinear distortion. This is quite similar to that in \cite{sebald2000support}.
	
	In the following sections, the nonlinear function in Eq. (\ref{eq:non_linear}) is considered, which is the same as \cite{salamanca2010channel,olmos2010joint,ye2017initial}:	
	\begin{equation}\label{eq:non_linear}
	|g(v)| = |v| + 0.2|v|^{2} - 0.1|v|^{3} + 0.5\cos(\pi|v|).
	\end{equation}
	
\begin{figure}[!h]
		\centering
		\begin{tikzpicture}[font=\sffamily\footnotesize]		
		\begin{semilogyaxis}[
		height=0.32\textheight,
		width=\linewidth,
		xmin=2, xmax=7,
		ymin=1e-2,
		grid=both,
		xtick = {2,3,4,5,6,7},
		xlabel=$E_{b}/N_{0}$ (dB),
		ylabel=BER,
		y label style={at={(axis description cs:-0.1,.5)}},
		line width=0.8pt,
		legend pos=south west
		]

		\addplot[mycolor2, mark=triangle, mark options={scale=0.8, solid}] coordinates {
			(2, 0.20558 )
			(3, 0.17150 )
			(4, 0.14016 )
			(5, 0.12080 )
			(6, 0.08734 )
			(7, 0.060187 )
		};
		\addlegendentry{SVM}
		
		\addplot[mycolor3, mark=o, mark options={scale=0.8, solid}] coordinates {
			(2, 0.16558 )
			(3, 0.1405 )
			(4, 0.11691 )
			(5, 0.09204 )
			(6, 0.07087 )
			(7, 0.050673 )
		};
		\addlegendentry{GPC}		
		
		\addplot[mycolor1, mark=diamond, mark options={scale=0.8, solid}] coordinates {
			(2, 0.1561315 )
			(3, 0.1290034 )
			(4, 0.1029875 )
			(5, 0.0791171 )
			(6, 0.0586809 )
			(7, 0.0414518 )
		};
		\addlegendentry{CNN}

		\end{semilogyaxis}
		\end{tikzpicture}
		\caption{BER comparison of proposed CNN, SVM and GPC \cite{olmos2010joint} equalizers over nonlinear channel.}
		\label{fig:ber_nonl_channel}
	\end{figure}
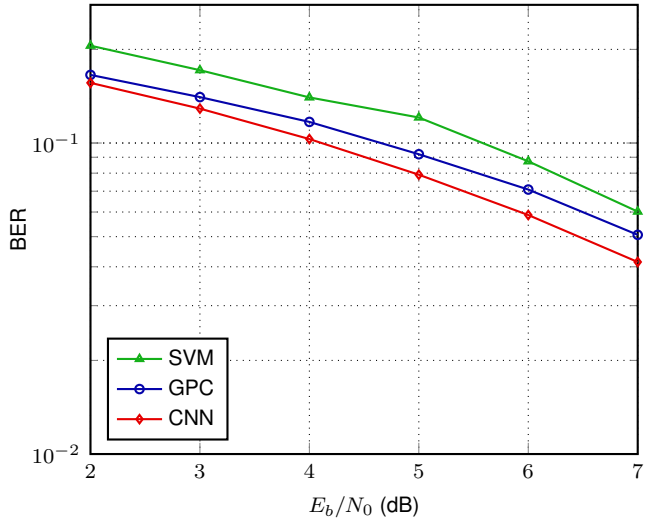

	For the case of nonlinear channel, BCJR method requires accurate CSI, and its computational complexity grows exponentially with the code length. Machine learning methods based equalization schemes show promising performance and lower complexity. We compare proposed CNN equalizer with GPC \cite{olmos2010joint} and SVM algorithms over the nonlinear channel in Eq. \ref{eq:channel_nl} and ISI in Eq. \ref{eq:ISI_H(z)}. Consistent with \cite{olmos2010joint}, a preamble sequence with 200 training samples is used for channel estimation. As illustrated Fig. \ref{fig:ber_nonl_channel}, CNN outperforms GPC and SVM with approximately 0.2 dB to 0.5 dB gain in terms of BER.
	
	\subsection{Experiment Results of Joint Equalizer and Decoder}
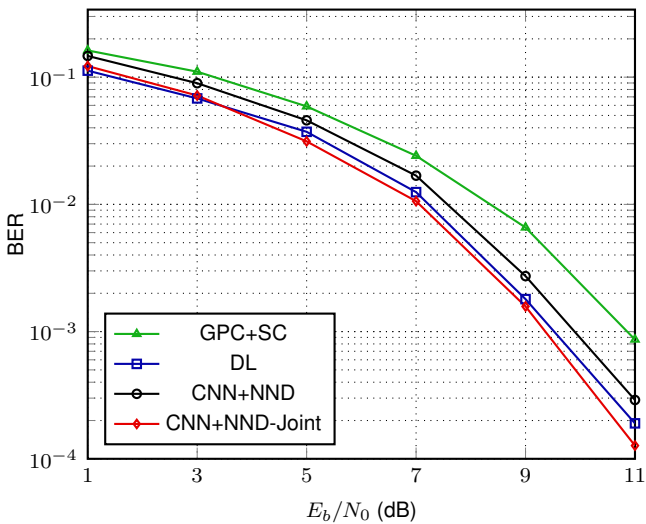
\begin{figure}[ht]
		\centering
		\begin{tikzpicture}[font=\sffamily\footnotesize]		
		\begin{semilogyaxis}[
		height=0.32\textheight,
		width=\linewidth,
		xmin=1, xmax=11,
		ymin=1e-4,
		grid=both,
		xtick = {1,3,5,7,9,11},
		xlabel=$E_{b}/N_{0}$ (dB),
		ylabel=BER,
		y label style={at={(axis description cs:-0.1,.5)}},
		line width=0.8pt,
		legend pos=south west
		]
		
		\addplot[mycolor2, mark=triangle, mark options={scale=0.8, solid}] coordinates {
			(1,		0.162180 )
			(3,		0.110410 )
			(5,		0.058930 )
			(7,		0.024092 )
			(9,		0.006581 )
			(11,	0.000866 )
		};
		\addlegendentry{GPC+SC}
		
		\addplot[mycolor3, mark=square, mark options={scale=0.8, solid}] coordinates {
			(1,		0.11249 )
			(3,		0.06807 )
			(5,		0.03719 )
			(7,		0.01248 )
			(9,		0.00180 )
			(11,	0.00019 )
		};
		\addlegendentry{DL}		
		
		\addplot[mark=o, mark options={scale=0.8, solid}] coordinates {
			(1,		0.14684 )
			(3,		0.08967 )
			(5,		0.04579 )
			(7,		0.01682 )
			(9,		0.00273 )
			(11,	0.00029 )
		};
		\addlegendentry{CNN+NND}
		
		\addplot[mycolor1, mark=diamond, mark options={scale=0.8, solid}] coordinates {
			(1,		0.122202 )
			(3,		0.071717 )
			(5,		0.031282 )
			(7,		0.010595 )
			(9,		0.001577 )
			(11,	0.000127 )
		};
		\addlegendentry{CNN+NND-Joint}
		
		\end{semilogyaxis}
		\end{tikzpicture}
		\caption{BER performance of various joint equalization and decoding scheme over nonlinear channel. GPC+SC and DL methods are redrawn from \cite{ye2017initial}.}
		\label{fig:ber_cnn_nnd}
	\end{figure}

	In the third experiment, a DNN with structure $\{16,128,64,32,8\}$ is introduced as the NND to increase the error correcting capability. The NND receives the soft output of CNN equalizer to recover the originally transmitted bits. The CNN trained over nonlinear channel in Eq. (\ref{eq:non_linear}) with ISI in Eq. (\ref{eq:ISI_H(z)}) and the NND trained over AWGN channel can be cascaded directly (CNN+NND). But the CNN output may not necessarily follow the same distribution of AWGN channel output, which will potentially cause degradation. Hence, joint training NND using the soft output of CNN equalizer (CNN+NND-Joint) can compensate the performance loss. We test on polar code (16,8) and the baselines are identical with \cite{ye2017initial}. As shown in Fig. \ref{fig:ber_cnn_nnd}, GPC+SC denotes the GPC equalization and successive cancellation (SC) decoding while DL denotes the deep learning method in \cite{ye2017initial}. CNN+DNN outperforms GPC+SC with 1 dB, and joint training of CNN and NND has about 0.5 dB gain over CNN+DNN. Performance of CNN+DNN-Joint scheme is very close to DL method, and there is a slight improvement on high SNR. The advantages of proposed CNN+NND is that it requires much less parameters ($\approx 15,000$) than DL \cite{ye2017initial} ($\approx 48,000$) without increasing the complexity. Moreover, CNN equalizer is also feasible for long codes.
	
	\section{Analysis}\label{sec:Analysis}
	\subsection{Related Work}
	CNN is widely used in image denoising \cite{jain2009natural}. Inspired by this, a CNN is cascaded behind the belief propagation (BP) decoder to deal with correlated Gaussian noise in \cite{liang2018iterative}. However, the limitation is that conventional BP decoder is untrainable thus unable to cope with ISI and nonlinear distortion, which may cause degradation to the whole system. Our proposed CNN equalizer learns the features directly from channel output, providing soft messages to the following channel decoder. This is more robust due to the strong adaptability of neural network. Compared with the DNN equalizer in \cite{ye2017initial}, CNN equalizer trained under short codes can be easily extended to arbitrary code length with negligible degradation. And we consider both ISI and nonlinearities rather than just ISI in \cite{caciularu2018blind}. The system architecture is similar to the CNN based blind detection scheme for SCMA in \cite{yang2018blind}. More complex channel impairments are taken into consideration in this paper.
	
	\subsection{Complexity Analysis}
	The complexity of CNN forward inference is determined by filter size $K$, input length $n$ and network depth $D$. Proposed CNN equalizer performs 1-D convolution, associating complexity $\mathcal{O}(n)$. Once trained, filter size and network depth are fixed. Hence, the overall complexity of CNN equalizer is $\mathcal{O}(n)$, indicating that the computational complexity grows linearly with sequence length. According to \cite{olmos2010joint}, the inference complexity of SVM is $\mathcal{O}(n)$ while GPC requires $\mathcal{O}(n^{2})$ without any optimization.

	\section{Conclusion}\label{sec:Conclusion}
	We present a novel CNN based equalizer to reduce the effects of channel impairments. Then a polar NND is utilized to recover the correct information. The CNN equalizer is adaptive for different channel conditions and shows substantial improvements compared with state-of-the-art results. The future work will focus on the efficient implementation of CNN equalizer and explore the possibility of joint equalization and decoding over long codes based on \cite{nachmani2018deep} and \cite{xu2017improved}.

	\footnotesize
	\bibliographystyle{IEEEtran}
	\bibliography{Bib}

\begin{thebibliography}{10}
\providecommand{\url}[1]{#1}
\csname url@samestyle\endcsname
\providecommand{\newblock}{\relax}
\providecommand{\bibinfo}[2]{#2}
\providecommand{\BIBentrySTDinterwordspacing}{\spaceskip=0pt\relax}
\providecommand{\BIBentryALTinterwordstretchfactor}{4}
\providecommand{\BIBentryALTinterwordspacing}{\spaceskip=\fontdimen2\font plus
\BIBentryALTinterwordstretchfactor\fontdimen3\font minus
  \fontdimen4\font\relax}
\providecommand{\BIBforeignlanguage}[2]{{%
\expandafter\ifx\csname l@#1\endcsname\relax
\typeout{** WARNING: IEEEtran.bst: No hyphenation pattern has been}%
\typeout{** loaded for the language `#1'. Using the pattern for}%
\typeout{** the default language instead.}%
\else
\language=\csname l@#1\endcsname
\fi
#2}}
\providecommand{\BIBdecl}{\relax}
\BIBdecl

\bibitem{dorner2018deep}
S.~D{\"o}rner, S.~Cammerer, J.~Hoydis, and S.~ten Brink, ``Deep learning based
  communication over the air,'' \emph{IEEE Journal of Selected Topics in Signal
  Processing}, vol.~12, no.~1, pp. 132--143, 2018.

\bibitem{nachmani2018deep}
E.~Nachmani, E.~Marciano, L.~Lugosch, W.~J. Gross, D.~Burshtein, and Y.~Be'ery,
  ``Deep learning methods for improved decoding of linear codes,'' \emph{IEEE
  Journal of Selected Topics in Signal Processing}, vol.~12, no.~1, pp.
  119--131, 2018.

\bibitem{arikan2009channel}
E.~Arikan, ``Channel polarization: A method for constructing capacity-achieving
  codes for symmetric binary-input memoryless channels,'' \emph{IEEE
  Transactions on Information Theory}, vol.~55, no.~7, pp. 3051--3073, 2009.

\bibitem{gruber2017deep}
T.~Gruber, S.~Cammerer, J.~Hoydis, and S.~ten Brink, ``On deep learning-based
  channel decoding,'' in \emph{Annual Conference on Information Sciences and
  Systems (CISS)}.\hskip 1em plus 0.5em minus 0.4em\relax IEEE, 2017, pp. 1--6.

\bibitem{xu2017improved}
W.~Xu, Z.~Wu, Y.-L. Ueng, X.~You, and C.~Zhang, ``Improved polar decoder based
  on deep learning,'' in \emph{IEEE International Workshop on Signal Processing
  Systems (SiPS)}, 2017, pp. 1--6.

\bibitem{ye2017initial}
H.~Ye and G.~Y. Li, ``Initial results on deep learning for joint channel
  equalization and decoding,'' in \emph{IEEE Vehicular Technology Conference
  (VTC-Fall)}, 2017, pp. 1--5.

\bibitem{caciularu2018blind}
A.~Caciularu and D.~Burshtein, ``Blind channel equalization using variational
  autoencoders,'' \emph{arXiv preprint arXiv:1803.01526}, 2018.

\bibitem{olmos2010joint}
P.~M. Olmos, J.~J. Murillo-Fuentes, and F.~P{\'e}rez-Cruz, ``Joint nonlinear
  channel equalization and soft ldpc decoding with gaussian processes,''
  \emph{IEEE Transactions on Signal Processing}, vol.~58, no.~3, pp.
  1183--1192, 2010.

\bibitem{salamanca2010channel}
L.~Salamanca, J.~J. Murillo-Fuentes, and F.~P{\'e}rez-Cruz, ``Channel decoding
  with a bayesian equalizer,'' in \emph{IEEE International Symposium on
  Information Theory Proceedings (ISIT)}, 2010, pp. 1998--2002.

\bibitem{kechriotis1994using}
G.~Kechriotis, E.~Zervas, and E.~S. Manolakos, ``Using recurrent neural
  networks for adaptive communication channel equalization,'' \emph{IEEE
  transactions on Neural Networks}, vol.~5, no.~2, pp. 267--278, 1994.

\bibitem{kingma2014adam}
D.~P. Kingma and J.~Ba, ``Adam: A method for stochastic optimization,''
  \emph{arXiv preprint arXiv:1412.6980}, 2014.

\bibitem{sebald2000support}
D.~J. Sebald and J.~A. Bucklew, ``Support vector machine techniques for
  nonlinear equalization,'' \emph{IEEE Transactions on Signal Processing},
  vol.~48, no.~11, pp. 3217--3226, 2000.

\bibitem{jain2009natural}
V.~Jain and S.~Seung, ``Natural image denoising with convolutional networks,''
  in \emph{Advances in Neural Information Processing Systems}, 2009, pp.
  769--776.

\bibitem{liang2018iterative}
F.~Liang, C.~Shen, and F.~Wu, ``An iterative bp-cnn architecture for channel
  decoding,'' \emph{IEEE Journal of Selected Topics in Signal Processing},
  vol.~12, no.~1, pp. 144--159, 2018.

\bibitem{yang2018blind}
C.~Yang, W.~Xu, X.~You, and C.~Zhang, ``A blind detection for {SCMA} based on
  image processing techniques,'' in \emph{IEEE International Symposium on
  Circuits and Systems (ISCAS)}, 2018.

\end{thebibliography}
	
\end{document}